\documentclass[pra,aps,groupedaddress,twocolumn,floatfix,nofootinbib]{revtex4-2}

\newcommand{\beq}{\begin{equation}}
\newcommand{\eeq}{\end{equation}}
\newcommand{\beqa}{\begin{eqnarray}}
\newcommand{\eeqa}{\end{eqnarray}}

\def\opone{\leavevmode\hbox{\small1\normalsize\kern-.33em1}}

\begin{document}

\title{The Multiverse Pandemic}
\author{Nicolas Gisin}
\affiliation{Group of Applied Physics, University of Geneva, 1211 Geneva 4, Switzerland\\
Schaffhausen Institute of Technology, SIT.org, Geneva, Switzerland}

\date{\small \today}
\begin{abstract}
I argue against the many-world interpretation (MWI) of quantum theory by emphasizing that when everything is entangled with everything else, in one big monstrous piece, there is no room left for creativity. Since the MWI was invented, it proves itself wrong (appeared first in French in 2010\cite{GdHasard}).
\end{abstract}
\maketitle

Like a disease, the multiverse is spreading. All sorts of communities have been infected by the multiverse\footnote{In French multi-vers means many worms.}, the many-worlds interpretation of quantum theory\cite{MWIVaidman}. Fortunately, as with other devastating pandemics, there are ways to protect oneself. Hence, I decided (hey yes, I enjoy the faculty of making decisions!) to write this short note for those who love life.

Remember Laplace: for a vast enough intellect, the future, just like the past, is fully determined by the present. In those bad old days, life was hard. The dictator Determinism reigned supreme; Newton’s laws were implacable (and his gravitation nonlocal!). There was no room for spontaneous phenomena, and no non-programmed events were tolerated. 

Nevertheless, many people survived, including those gifted with free-will. How could they? Well, thanks to Descartes they knew that their free-will could express itself by spontaneous forces acting on the material world. Admittedly, the interface between their will and the material world was quite elusive, but Descartes gave it a name: the pineal gland. This was just a name, but a very important name: naming this interface demonstrated that the existence of free-will is not in contradiction with deterministic classical physics. Newton never pretended that his physics were complete.

And so, the dictatorship of Determinism was tolerable to free men.

Then came quantum physics. At first, free men celebrated the revolution of intrinsic randomness in the material world. This was the end of the awful dictator Determinism, or so they thought. But this dictator had a son… or was it his grandson?

Determinism returned in the new guise of quantum physics without randomness: everything, absolutely everything, all alternatives, would equally happen, all on an equal footing. Real choices were no longer possible. But the most terrible was still to come: universal entanglement. According to the new multiversal dictator, not only did the material world obey deterministic laws, but it was all one big monstrous piece, everything entangled with everything else. There was no room left for any pineal gland, no possible interface between physics and free-will. The sources of all forces, all fields, everything was part of the big $\Psi$, the wavefunction of the multiverse, as the dictator bade people call their new God.

But fortunately, the son (or grandson) or the former dictator was not as strong as his ancestor. Quite a few physicists adopted another religion, a less demanding one, whose mantra was “shut-up and calculate”. A schism occurred, but an abundance of new results in quantum physics allowed the different sects to live in peaceful coexistence… at least for a while. 

But “shut-up and calculate” is not a very attractive creed. And what was feared came to pass: the multiverse pandemic spread, reaching first the weakest, many young physicists were infected. The arguments of the dictator’s priests were simple, and hence efficient: “our religion is the simplest, hence it must be true”. And for the sceptic they added, “if you don’t believe in our dictator, you’ll be cut down by Ockham’s Razor”. What? Ockham’s Razor would favour the multiverse? Yes, claimed the priests, because by rejecting the many-worlds, you commit the crime of modifying the Schrödinger equation. Adding corrections to the celebrated Schrödinger equation is worse than adding worlds, claimed the priests\cite{Zeh}.

The argument seemed strong, and the pandemic spread and spread. Terrifyingly, not only was the reign of Determinism back, but without the small alcove where the pineal gland used to sit.

It’s time to take a step back. I am a free being, I enjoy free-will. I know that much more than anything else. How then, could an equation, even a truly beautiful equation, tell me I’m wrong? I know that I am free much more intimately than I will ever know any equation. Hence, and despite the grandiloquent speeches, I know in my gut that the Schrödinger equation can’t be the full story; there must be something else. “But what?”, reply the dictator’s priests. Admittedly, I don’t know, but I know the multiverse hypothesis is wrong, simply because I know determinism is a sham\cite{Qrandom,GargantuanChaos,SpontaneousStochasticity,FlavioNGPRA19,RealNbs,NatureComment}.

\end{document}